# On the possible mechanisms of the selective effect of a non-equilibrium plasma on healthy and cancer cells in a physiological solution

**Mikhail N. Shneider [1*], Mikhail Pekker**

[1]Department of Mechanical and Aerospace Engineering, Princeton University, Princeton, NJ USA
[*]m.n.shneider@gmail.com

**Abstract**

This paper discusses possible mechanisms for the selective effect of weakly ionized nonequilibrium plasma and currents in electrolyte on healthy and cancerous cells in physiological saline in a Petri dish. The interaction with the plasma source leads to a change in osmotic pressure, which affects the electro-mechanical properties of cell membranes in healthy and cancerous cells in different ways. The currents arising in the electrolyte charge the membranes of healthy and cancerous cells to a different potential difference due to the different values of the membranes' dielectric constant. We hypothesized that the dielectric permeability of cancer cell membranes is lower than that of healthy cells, as is the capacity of a unit of the membrane surface, and therefore, the additional potential difference acquired by the membrane through charging with currents induced in the intercellular electrolyte is greater in cancer cells. This can lead to electroporation of cancer cell membranes, resulting in their apoptosis, but does not affect healthy cells.

A number of experimental studies have convincingly shown that the action of weakly ionized nonequilibrium plasma on a physiological solution can selectively affect the cells present in that solution [1-9]. Special attention should be paid to works in which the selective stimulation of apoptosis in cancer cells was observed [4-9].

In [10], we described the importance of accounting for osmotic pressure when analyzing the physiological effects on cellular structures in plasma medicine. The interaction of a weakly ionized plasma jet with a saline solution leads to detectable changes in the saline's ion-molecular composition and, hence, changes in the osmotic pressure drop across the cell membranes. In turn, changes in osmotic pressure lead to water flow through the membrane and, correspondingly, to cell compression. Estimations of the compression coefficient for the membranes of some types of cells are presented in [16].

It was shown in [10] that the interaction of weakly ionized plasma with saline leads to the diffusive flow of hydrated ions from the interface to the cell structures, thus changing the ionic composition of the saline electrolyte and, therefore, the osmotic pressure drop across the cell membranes (Fig. 1(A)) [6].

It is quite obvious that membrane compression could change the membrane's porosity and, correspondingly, its dielectric permissibility due to the density change of the water molecule in it and the ion and protein transport through it. However, if the plasma interacting with the electrolyte has a potential different from that of the floating equilibrium in the electrolyte, then ion currents are induced in the electrolyte. Currents can be excited without plasma when electrodes are immersed in the electrolyte, to which a potential difference is applied (Fig. 1(B)) [11-15]. These ion currents can lead to an additional charge of cell membranes, thus changing the transmembrane potential difference. It is also possible that the combined effect of changes in the ionic composition resulting from the interaction of the plasma with the solution and from the currents induced in the Petri dish (Fig. 1(C)) is due to the fact that the region of the interaction of the electrolyte with the plasma, in general, has a potential different from the floating potential in the electrolyte.

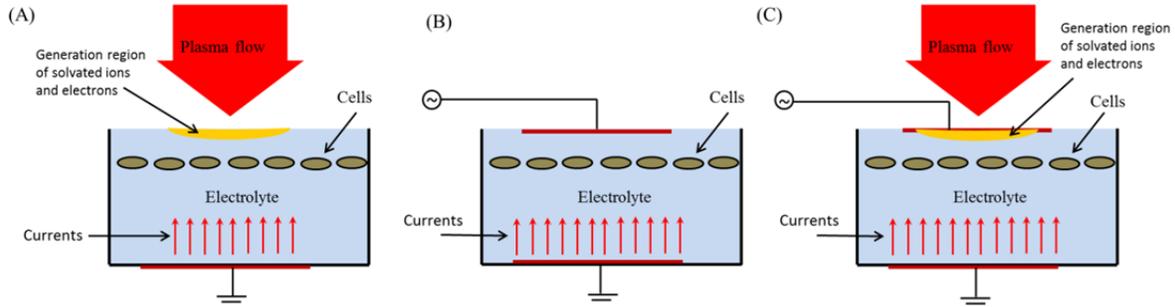

**Fig. 1.** Schematics of the typical experimental setups with cell structures in saline in Petri dish. (A) - plasma jet interacting with air-saline interface. (B) – currents induced by voltage applied to the electrodes. (C) – joint effect of applied voltage and plasma jet.

Membrane properties in cancer and healthy cells can vary greatly. Therefore, the module of the all-round compression and shear modulus for hepatocellular carcinoma cells (HCCs) in the measurements [16] are equal to $K_1 = 103.6 \text{ Nm}^{-2}$ and $K_2 = 42.5 \text{ Nm}^{-2}$, respectively, and for hepatocytes, they are $K_1 = 87.5 \text{ Nm}^{-2}$ and $K_2 = 33.3 \text{ Nm}^{-2}$, respectively. The measured viscosity coefficients in the membrane also are different: $\mu = 4.5 \text{ Pa} \cdot \text{s}$ in HCCs and $\mu = 5.9 \text{ Pa} \cdot \text{s}$ in rat hepatoma cells. Resting potentials on the membranes of healthy and cancerous cells also differ markedly. For example, [17] showed that the transmembrane resting potentials in rat hepatocytes and hepatoma cells are -37.1 and -19.8 mV, respectively, and in mouse corneal fibroblast and fibrosarcoma cells, the corresponding resting potentials are -42.5 and -14.3 mV. In other words, the resting potential in cancer cells is two times less than in healthy ones, indicating a difference in the transport properties of the membranes of healthy and cancerous cells. Therefore, even a small effect on the mechanical properties of the membranes or on the transmembrane potential difference can have a significant effect on the transport into and out of the healthy and cancer cells.

The currents arising in the electrolyte charge the membranes of healthy and cancerous cells to a different potential difference due to the different values of the membranes' dielectric constant. One of the paper's hypotheses is that the dielectric permeability of cancer cell membranes is lower than that of healthy cells, as is the capacity of a unit of membrane surface, and, therefore, the additional potential difference acquired by the membrane through charging with currents induced in the intercellular electrolyte is greater in cancer cells. This can lead to electroporation of cancer cell membranes, resulting in their apoptosis, and does not affect healthy cells.

In this brief paper, we would like to bring the scientific community's attention to the fact that change in the cell membrane surface charge due to the currents in electrolyte may be responsible for selective action on healthy and cancer cells due to, as mentioned above, the difference in the mechanical-electrical properties of cell membranes and the different responses to the change in the membrane charge. In the electrolyte, currents can be produced by an electrode immersed in a Petri dish or by plasma jet (Fig.1). It should be noted that, in the case of a plasma jet and electrodes (Fig. 1(C)), the membrane charging will be accompanied by a change in osmotic pressure, which enhances the effect on the cells.

The application of electric fields to biological cells in a conducting medium leads to an accumulation charge on the outer surface of the membrane and, consequently, to a change in voltage across the membrane. Small changes in the potential on the membrane (~10–100 mV) of short duration in the order of milliseconds lead to a disruption of the voltage-dependent channels and, accordingly, to a change in the ion transport through the membrane without destroying the integrity of the membrane [11-13].

With higher electric fields related to a correspondingly higher voltage across the cell membrane, electroporation occurs. Thus, the permeability of the membrane increases to such a level that either the cell needs from seconds to hours to recover (reversible breakdown) or cell death occurs (irreversible breakdown) [11-15]. In [11-13], it was experimentally shown that the critical field strength for the lysing of bacteria (prokaryotic cells) is around 0.1–1 V across the cell membrane, depending on exposure time. In [15], experimental results were given for the influence of electric fields of varying intensity and duration on cancer cells at different values for the medium's conductivity [15]. It was shown that, as the medium's salinity (conductivity) increases, the electric field at which cell death occurs begins to fall.

For simplicity, we will assume that the cells are spherical. Consider a cell of radius $a$ immersed in an electrolyte in which a current is maintained by an external source of electromotive force. The continuity equation for the currents flowing in the extracellular electrolyte is

$$div\vec{j} = div(\sigma\vec{E}) = \sigma div(\vec{E}) = -\sigma\Delta\varphi = -\sigma\left(\frac{1}{r^2}\frac{\partial}{\partial r}r^2\frac{\partial\varphi}{\partial r} + \frac{1}{r^2\sin\theta}\frac{\partial}{\partial\theta}\sin\theta\frac{\partial\varphi}{\partial\theta}\right) = 0, \qquad (1)$$

where $\varphi, \vec{j}, \text{and } \vec{E}$ are distributions of potential, currents, and electric field and $\sigma$ is the electrolyte conductivity.

In (1) we used Ohm's law $\vec{j} = \sigma\vec{E}$, and the relation $\vec{E} = -\nabla\varphi$.

Assuming that the membrane is impermeable to ion currents, as a result of charging the membrane, the stationary radial electric field must be zero:

$$E_n = -\frac{\partial\varphi}{\partial r}\bigg|_{r=a} = 0. \qquad (2)$$

At infinity

$$j_z|_{z=\pm\infty} = j_0 = \sigma E_0. \qquad (3)$$

The solutions for (1) with boundary conditions (2, 3) are

$$\varphi_{r\geq a} = E_0 r \cos\theta\left(1 + \frac{a^3}{2r^3}\right), \qquad (4)$$

and, at $r = a$,

$$\varphi_a = \frac{3}{2}E_0 a \cos\theta = \frac{3}{2\sigma}j_0 a \cos\theta. \qquad (5)$$

From (5), it follows that $\varphi_a$ in (6) is different at $\theta = 0$ and $\theta = \pi$. Accordingly, the additional potential differences induced by currents on the membrane are as follows:

$$\varphi_a(0) = \frac{3}{2}E_0 a, \text{ at } \theta = 0, \qquad (6a)$$

and

$$\varphi_a(\pi) = -\frac{3}{2}E_0 a, \text{ at } \theta = \pi. \tag{6b}$$

Since the potential difference across the membrane in cancer cells is twice as large as that of healthy cells [17], a potential change of ~100 mV can cause irreversible breakdown of the membrane of cancer cells like hepatoma and fibrosarcoma cells but can be safe for healthy cells like hepatoma and fibroblast cells.

Let us find the value $j_0$ at which the additional change of the potential difference on the membrane due to being charged by currents in the electrolyte is $\varphi_a$~100 mV. Therefore, for example, for $a \approx 11$ μm, corresponding to the size of the hepatocytes of SH-R3s [18] and $\sigma = 1$ S/m [19], we obtain

$$j_0 \approx \frac{2\sigma}{3a}|\varphi_a| = 6 \cdot 10^3 \text{ A/m}^2. \tag{7}$$

Then, let us estimate the charging time of the surface of the sphere to the potential (6). Obviously, this time is determined by the time required to charge the spherical capacitor to the potentials from (6a) and (6b). For simplicity, we shall assume that a thin-walled dielectric sphere does not perturb the electric field near its surface. In this case, the radial current charging the spherical capacitor is

$$j_n = -j_0 cos(\theta). \tag{8}$$

In this case, the estimate of the transition time (characteristic charging time of the capacitor) is

$$\tau_{ch} \sim c_m \frac{|\varphi_a|}{j_n} = \frac{3}{2}\frac{c_m a}{\sigma} \sim 0.15 \text{ μs}. \tag{9}$$

In (9), we substituted $c_m \approx 10^{-2}$ F/m$^2$ [19,20], $a = 10$ μm, and $\sigma = 1$ S/m [19]. If the change of fields in the electrolyte surrounding the cell lasts more than a few tenths of a microsecond, the induced potential on the membrane can be considered quasi-permanent, as defined by (5). If the current "at infinity" changes its direction for a time much shorter than the membrane charging time (9), the change in the membrane potential caused by its additional charging does not affect the operation of the voltage-dependent channels and, accordingly, the cell activity [13]. Note that we have neglected Joule heating. This is a fully justified assumption since an elementary estimate shows that heating of saline with conductivity $\sigma = 1$ S/m up to $\Delta T = 1^0$ K by the current density (7) takes more than 0.1 second. This time is much longer than all the characteristic times of the problem under consideration, even if the heat conduction losses are neglected.

It is known that, when a nonequilibrium plasma jet [21] or other types of weakly ionized plasma, for example, a dielectric barrier discharge plasma [1,22], interacts with a physiological solution, long-lived solvated (hydrated) ions are formed in the interface. It was shown in [10] that the diffusion of these ions into the depth of the Petri dish, where the studied cell cultures are located, leads to an increase in the density of hydrated ions near the cells and, accordingly, to a change in the osmotic pressure since the membrane is permeable to water but impermeable to hydrated ions. As a result, it is most likely that a change in the concentration of hydrated ions in the intercellular medium makes the solution a hypertonic solution, which leads to the release of excess water from the cell and cell compression [10].

However, a change in the density of ions in the solution also leads to an increase in the conductivity of the medium and to an increase in the charging current of the membrane. The process of charging the membrane depends on the type of plasma source interacting with the boundary of saline and feeding hydrated ions into the solution. In the case of a Dielectric Barrier Discharge (DBD) plasma source with nanosecond repetitive pulses [22], the amount of time the change of the electric field lasts is much less than the charging time of the membrane $\tau_{ch}$ (9), and the induced currents obviously cannot noticeably affect the charge on the membrane. However, with continuous plasma exposure of a microsecond or longer, charging cell membranes with currents induced in the electrolyte, together with altered osmotic pressure, should certainly affect the mechanical properties of the membranes and the transmembrane transport through ion and other channels.

The dielectric constant of the real phospholipid membrane in the cell, estimated by the experimental value of the capacitance of the membrane, is of the order of $\varepsilon_m \sim 7 - 10$ [19,20], while in the ideal phospholipid membrane, the relative dielectric permittivity is $\varepsilon_m \sim 2 - 3$ [23]. The difference is two- to threefold, apparently due to the presence of water molecules in the pores and defects of the real cell membrane. Since the compression of the membrane can result in its dehydration, the change in osmotic pressure can lead to a change in the electrical properties of the membrane (i.e., an additional potential difference acquired when charging the membrane) up to a loss of dielectric strength (electroporation). Furthermore, if the electrical and mechanical properties of cell membranes in healthy and cancer cells are different, then the induced currents in saline and the osmotic pressure changes cause different physiological consequences, resulting in the selectivity of the plasma effects on cell structures observed in the experiments.

As we saw above with examples of hepatocellular carcinoma (HCC) and hepatocyte cells, the module of the all-round compression of cancer cells significantly exceeds the module of the all-round compression of the healthy cells [16]. Therefore it should be expected that, with the same change in osmotic pressure, the compression of the cancer cells and hence, the displacement of water from it, are less significant than in healthy cells. This means that the change in volume of a cancer cell caused by an additional osmotic pressure differential is less than a change in the volume of healthy cells. Thus, after the interaction of plasma with saline, the final osmotic pressure drop on the cancer cells membranes is significantly greater than on the membranes of healthy cells. This leads to a greater displacement of water from the membrane of cancer cells than from healthy cells. As a result, the dielectric constant and the specific capacity of cancer cell membranes become significantly smaller than those of healthy cell membranes. Thus, with the same current and exposure time, the membranes of cancer cells are charged to a significantly higher potential difference, leading to a more likely electroporation. This may cause apoptosis in cancer cell, but not in healthy cell membranes. This may explain the selectivity of apoptosis of cell cultures in a Petri dish when interacting with plasma.

Measuring the dielectric permittivity of the membranes of healthy and cancerous cells is very important for answering the question about the possible mechanism of the selective effect of weakly ionized plasma and the currents it causes on the cell structures in a physiological solution.

**Literature**
[1] A.A. Fridman, G.D. Friedman, Plasma Medicine (NY John Wiley and Sons Ltd, 2013)
[2] M. Laroussi, J.P. Richardson, F.C. Dobbs, Effects of Non-Equilibrium Atmospheric Pressure Plasmas on the Heterotrophic Pathways of Bacteria and on their Cell Morphology, Appl. Phys. Lett. **81**, 772, (2002)
[3] X. Lu, G.V. Naidis, M. Laroussi, S. Reuter, D.B. Graves, K. Ostrikov, Reactive species in non-equilibrium atmospheric-pressure plasmas: generation, transport, and biological effects. Physics Reports. **630**, 1 (2016)
[4] G.J. Kim, W. Kim, K.T. Kim, J.K. Lee, DNA damage and mitochondria dysfunction in cell apoptosis induced by nonthermal air plasma. Applied Physics Letters, **96**(2), 021502 (2010)
[5] N. Barekzi, M. Laroussi, Dose-dependent killing of leukemia cells by low-temperature plasma. J.Phys. D: Applied Physics **45**(42), 422002 (2012)
[6] D. Yan, J.H. Sherman, M. Keidar, Cold atmospheric plasma, a novel promising anti-cancer treatment modality, Oncotarget, **8**, 15977 (2017)

[7] D. Yan, H. Cui, W. Zhu, N. Nourmohammadi, J. Milberg, L.G. Zhang, J. H. Sherman, M. Keidar, The specific vulnerabilities of cancer cells to the cold atmospheric plasma-stimulated solutions, Scientific Reports **7**, 4479 (2017)
[8] D. Yan, A. Talbot, N. Nourmohammadi, J. Sherman, X. Cheng, M. Keidar, Toward understanding the selectivianti-cancer capacity of cold atmospheric plasma - a model based on aquaporins, Biointephases, **10**, 040801 (2015)
[9] D. Yan, H. Xiao, W. Zhu, N. Nourmohammadi, G. Zhang, K. Bian, M. Keidar, The role of aquaporins in the anti-glioblastoma capacity of the cold plasma-stimulated medium, J. Phys. D: Appl. Phys. **50**, 055401 (2017)
[10] M.N. Shneider, M. Pekker, Effect of weakly ionized plasma on osmotic pressure on cell membranes in a saline, Journal of Applied Physics **123**, 204701 (2018)
[11] K.H. Schoenbach, F.E. Peterkin, R.W. Alden, S.J. Beebe, The Effect of Pulsed Electric Fields on Biological Cells: Experiments and Applications, IEEE Trans. Plas. Sci., **25**, 284 (1997)
[12] U. Zimmermann, G. Pilwat, F. Beckers, F. Riemann, Effects of external electrical fields on cell membranes, Bioelectrochemistry and Bioenergetics, **3**, 58 (1976)
[13] H. Huelsheger, J. Potel, and E.G. Niemann, Killing of bacteria with electric pulses of high electric field strength, Radiat. Environ. Biophys. **20**, 53 (1981)
[14] K.H. Schoenbach, A. Abou-Ghazala, T. Vithoulkas, R.W. Alden, R. Tumer, S. Beebe, The effect of pulsed electrical fields on biological cells, 0 Digest of Technical Papers. 11th IEEE International Pulsed Power Conference (Cat. No.97CH36127) (1997)
[15] A. Silve, I. Leray, C. Poignard, L.M. Mir, Impact of external medium conductivity on cell membrane electropermeabilization by microsecond and nanosecond electric pulses, Scientific Reports, **6**, 19957 (2016)
[16] G. Zhang, M. Long, Z-Z Wu, and W-Q Yu, Mechanical properties of hepatocellular carcinoma cells, World J. Gastroenterol. **8,** 243 (2002)
[17] R. Binggeli, I.L. Cameron  Cellular, Potentials of Normal and Cancerous Fibroblasts and Hepatocytes, Cancer Research, **40**, 1830 (1980)
[18] S. Katayama, C. Tateno, T. Asahara, K. Yoshizato, Size-Dependent in Vivo Growth Potential of Adult Rat Hepatocytes, Am. J. Pathol. **158**, 97 (2001)
[19] A.L. Hodgkin, A.F. Huxley, A quantitative description of membrane current and its application to conduction and excitation in nerve, J. Physiol. **117**, 500 (1952)
[20] L. Goldman, J. S. Albus, Computation of Impulse Conduction in Myelinated Fibers; Theoretical Basis of the Velocity-Diameter Relation, Biophys. J. **8**, 596 (1968).
[21] P. Bruggeman, C. Leys, Non-thermal plasmas in and in contact with liquids, J. Phys. D: Appl. Phys. **42**, 053001 (2009)
[22] H. Ayan, D. Staack, G. Fridman, A. Gutsol, Y. Mukhin, A. Starikovskii, A. Fridman, G. Friedman, Application of nanosecond-pulsed dielectric barrier discharge for biomedical treatment of topographically non-uniform surfaces, J. Phys. D: Appl. Phys. **42**, 125202 (2009)
[23] M.V. Volkenstein, General Biophysics (Elsevier Science & Technology Books 1983)